\documentclass[12pt]{article}
\usepackage{epsfig}
\usepackage{graphicx}

\newsavebox{\sboxpubnumber}
\newsavebox{\sboxpubdate}
\newcommand{\pubdate}[1]{\begin{lrbox}{\sboxpubdate}{#1}\end{lrbox}}
\newcommand{\pubnumber}[1]{\begin{lrbox}{\sboxpubnumber}{\begin{tabular}{l} #1 \\
				 \usebox{\sboxpubdate}
				 \end{tabular}}
                           \end{lrbox}
                           \pubblock}
\newcommand{\Title}[1]{\begin{center} {\Large #1 } \end{center}}
\newcommand{\Author}[1]{\begin{center}{ \sc #1} \end{center}}
\newcommand{\Address}[1]{\begin{center}{ \it #1} \end{center}}

\newcommand{\pubblock}{\rightline{
			\usebox{\sboxpubnumber}}}
\newenvironment{Abstract}{\begin{quotation}  }{\end{quotation}}
\newenvironment{Presented}{\begin{quotation} \begin{center}
             PRESENTED AT\end{center}\bigskip
      \begin{center}\begin{large}}{\end{large}\end{center}
      \end{quotation}}
\newcommand{\Acknowledgements}{\bigskip  \bigskip \begin{center} \begin{large}
             \bf ACKNOWLEDGEMENTS \end{large}\end{center}}

\begin{document}

\begin{titlepage}
\pubdate{\today}
\pubnumber{UFIFT-HEP-01-21 \\ astro-ph/0111462} 

\vfill
\Title{Effective Field Equations of the Quantum \\ Gravitational Back-Reaction
on Inflation}
\vfill
\Author{R. P. Woodard\footnote{e-mail: woodard\@phys.ufl.edu.}}
\Address{Department of Physics, University of Florida \\
         Gainesville, FL 32611 USA}
\vfill

\begin{Abstract}
Quantum gravitational back-reaction offers the potential of simultaneously
resolving the problem of the cosmological constant and providing a natural
model of inflation in which scalars play no special role. In this model 
inflation begins because the cosmological constant is not unnaturally small. 
It ends through the accumulated gravitational interaction between virtual
gravitons which are ripped apart by the inflationary expansion. Although
perturbative techniques can be used to study the effect as long as it remains
weak, they break down when back-reaction begins to exert an appreciable effect
on the expansion rate. In this talk I argue that the end of inflation is 
sudden and that there is actually an overshoot into deflation. (This 
incidentally provides a very efficient mechanism for reheating.) The 
subsequent evolution can be understood in terms of a competition between the 
opening of the past light cone and the formation of a thermal barrier to the 
persistence of correlations from during the period of inflation.
\end{Abstract}
\vfill
\begin{Presented}
    COSMO-01 \\
    Rovaniemi, Finland, \\
    August 29 -- September 4, 2001
\end{Presented}
\vfill
\end{titlepage}
\def\thefootnote{\fnsymbol{footnote}}
\setcounter{footnote}{0}

\section{Introduction}

Quantum gravitational back-reaction offers an attractive model of cosmology
which also resolves the problem of the cosmological constant. The idea 
is that there is no fine tuning of the cosmological constant, $\Lambda$, 
or of scalar potentials \cite{TsWo1}. In fact there need not be any scalars. 
Inflation begins in the early universe because $\Lambda$ is positive and not 
unnaturally small. Inflation eventually ends due to the accumulation of 
gravitational attraction between long wavelength virtual gravitons which are 
pulled apart by the rapid expansion of spacetime. Inflation persists for many 
e-foldings because gravity is a weak interaction, even at typical inflationary 
scales, and it requires an enormous accumulation of gravitational potential to 
overcome this. Because the model has only a single free parameter --- $G 
\Lambda$, where $G$ is Newton's constant --- it can be used to make unique and 
testable predictions. For example, when $\Lambda$ is adjusted (to a mass scale
of about $.72 \times 10^{16}~{\rm GeV}$) so as to make the COBE 4-year RMS 
quadrupole\cite{COBE} come out right, the model gives the following predictions
for the parameters which describe primordial anisotropies in the cosmic 
microwave background \cite{ATW}:
\begin{equation}
r \approx .0017 \qquad , \qquad n_s \approx .97 \qquad , \qquad n_t \approx 
-.00028 \; .
\end{equation}

The mechanism through which long wavelength gravitons are pulled out of the
vacuum in an inflating universe is known as {\it superadiabatic amplification}
\cite{Grish}. It can be understood as the consequence of three well-known 
results. The first is that the energy-time uncertainty principle implies the 
continual emergence of all types of virtual quanta from empty space. The second 
fact is that inflation provides a geometry for which Zeno's argument against 
the possibility of motion is sometimes valid. Even light cannot pass between
co-moving observers which are separated by more than a Hubble length, $H^{-1}$,
where $H \equiv \sqrt{\Lambda/3}$. By the time the light has traveled halfway, 
the intervening space has expanded so much that the distance remaining is 
actually greater than at the beginning. It follows that virtual pairs of 
wavelength comparable to or greater than $H^{-1}$ cannot recombine. The final 
fact is that the graviton's masslessness and lack of conformal invariance 
imply an appreciable rate for the emission of virtual quanta of Hubble 
wavelength. It is significant that the analogous process for light, minimally 
coupled scalars is thought to be the origin of the observed anisotropies in 
the cosmic microwave background \cite{MFB,LL}. This is solid physics.

Particles which are massive at or above the scale of inflation do not 
experience superadiabatic amplification because the amplitude for producing 
virtual quanta of Hubble wavelength is so small. Nor is there any 
superadiabatic amplification of particles which are conformally invariant on 
the classical level. Their quanta cannot locally sense the expansion of
spacetime because the geometry of an inflating universe tends rapidly towards 
homogeneity and isotropy, and hence to conformal flatness. This means that the
graviton is almost unique. If a minimally coupled scalar can somehow avoid 
acquiring mass at or above the scale of inflation it can also experience 
superadiabatic amplification. All other particles are either massive or else 
they possess classical conformal invariance. If one was searching for an arena
in which quantum gravitational effects might be significant, the long 
wavelength sector of inflationary cosmology would be a natural choice.

The infrared character of the physical mechanism means that it can be studied
reliably using perturbative quantum general relativity, without requiring a 
fundamental theory of quantum gravity. Loops of massless particles give rise
to nonlocal and ultraviolet finite terms which cannot be subsumed into local
counterterms and which are not affected by changes in the short wavelength 
sector. Infrared phenomena can therefore be studied using the low energy 
effective theory. This is why Bloch and Nordsieck \cite{BN} were able to 
resolve the infrared problem in QED before the theory's renormalizability was 
suspected. It is also why Weinberg \cite{Wein} was able to give a similar 
resolution for the infrared problem of quantum gravity with zero cosmological 
constant. And it is why Feinberg and Sucher \cite{FS} were able to compute the 
long range force induced by neutrino exchange using Fermi theory. Extensive 
work along these same lines has been done recently on quantum gravity with 
zero cosmological constant by Donoghue \cite{Dono}.

Nick Tsamis and I have done a two loop computation of the pure quantum 
gravitational back-reaction on inflation \cite{TsWo2}. The expectation value of 
the gauge-fixed metric was computed on the manifold $T^3 \times R$ and in the 
presence of a state which is free graviton vacuum at $t=0$:
\begin{equation}
\Bigl\langle \Omega \Bigl\vert g_{\mu\nu}(t,{\vec x}) dx^{\mu} dx^{\nu} 
\Bigr\vert \Omega \Bigr\rangle = -dt^2 + e^{2 b(t)} d{\vec x} \cdot d{\vec x} 
\; .
\end{equation}
Treating this expectation value as if it were the classical invariant element, 
one forms the effective Hubble constant,
\begin{equation}
{\dot b}(t) = H \left\{1 - \left({G \Lambda \over 3 \pi}\right)^2 \left(
\frac16 (H t)^2 + O(H t)\right) + O(G^3)\right\} \; . \label{grresult}
\end{equation}
The correct interpretation of this result is that quantum gravitational 
back-reaction slows inflation as long as perturbation theory remains valid, and
that the effect must eventually grow nonperturbatively strong.

The great unsolved problem is how to analytically treat the period beyond the 
end of inflation when perturbation theory is no longer valid. That progress
is possible derives from the peculiar manner in which slowing becomes
nonperturbative. This is not a strongly coupled theory like QCD; the
dimensionless coupling constant is $G \Lambda \approx 5 \times 10^{-11}$. It 
is not that any particular particle creation event matters much but rather 
that an enormous number of them add coherently over the ever-increasing
volume of the past light cone. Those factors of $Ht$ in (\ref{grresult}) 
derive from the fact that there are two vertices being integrated over the
past light cone, whose invariant volume during inflation is,
\begin{equation}
V(t) = \int_0^t dt' e^{3 b(t')} \frac43 \pi \left[\int_{t'}^t dt''
e^{-b(t'')} \right]^3 \approx \frac43 \pi H^{-4} H t \; .
\end{equation}
{\it It is therefore reasonable to expect that the incremental screening 
from each volume element of the past light cone is always minuscule.} These
increments can be computed perturbatively if only we can learn how to work in 
the changing background.

My purpose here is to outline the physical principles and techniques by which 
Dr. Tsamis and myself are attempting to gain analytic control over the
nonperturbative regime. Section 2 describes the analog scalar model with which 
we test ideas. Section 3 is devoted to a new infrared expansion of the mode 
functions which allows us to obtain the propagators as the background changes. 
Section 4 argues that screening becomes so strong at the end of inflation that 
there is no alternative to a brief period of deflation. This leads to the rapid
formation of a hot, dense plasma of ultraviolet particles which we call, the 
{\it thermal barrier}. The thermal barrier scatters the infrared virtual quanta
which need to carry quantum correlations into the future from the period of 
inflationary particle production in order to keep the large bare cosmological 
constant screened. When the barrier becomes too effective, deflation ends and 
the universe begins to expand again. Section 5 discusses the balance between
the thermal barrier and the growth of the inflationary past light cone which
we believe governs this subsequent phase of expansion. Since we have not yet 
proven the viability of this scheme I can offer no definite conclusions. I 
close instead by mentioning the potential payoff if these ideas lead to a 
workable model.

\section{The scalar analog model}

As discussed in the previous section, the effect we seek to study has
two essential features. The first is that infrared virtual particles
are continually being ripped out of the vacuum and pulled apart by the
inflationary expansion of spacetime. The second crucial feature is that
these particles attract one another through a weak long range force which
gradually accumulates as more and more particles are created. The
particles we believe were actually responsible for stopping inflation
are gravitons, and the long range force through which they did it was
gravitation. However, this is not a simple theoretical setting in which
to work. It took over a year of labor to obtain the two loop result
(\ref{grresult}) --- even with computer symbolic manipulation programs
\cite{TsWo2}. Before attempting to test speculative ideas in quantum
general relativity it is natural to wonder if there is not a simpler 
theory which manifests the same effect.

As also explained in the previous section, the prerequisites for
inflationary particle production are masslessness on the scale of
inflation and the absence of classical conformal invariance. Massless,
minimally coupled scalars have these properties --- and the lowest
order back-reaction from self-interacting scalars can be worked out on a
blackboard in about 15 minutes \cite{TsWo3}. Of course it is not natural
for scalars to possess attractive self-interactions and still remain
massless on the scale of inflation. But we do not need a {\it realistic}
model --- that is already provided by gravitation. What we seek is rather
a {\it simple} model that can be tuned to show the same physics, however
contrived and unnatural this tuning may be.

\begin{figure}
\centerline{\epsfxsize=0.6\textwidth\epsffile{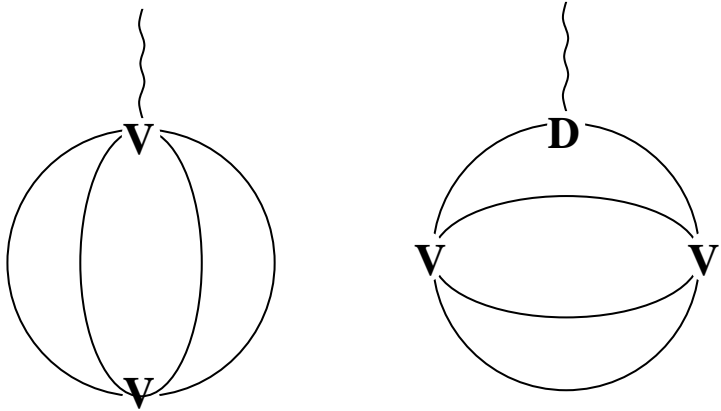}}
\caption{Contributions to the scalar stress-energy tensor at order $\lambda^2$
{\it with} covariant normal ordering. $V$ denotes the 4-point vertex and
$D$ represents the derivative vertex.}
\end{figure}

A massless, minimally coupled scalar with a nonderivative self-interaction
seems to do what we want,
\begin{equation}
{\cal L} = -\frac12 : \partial_{\mu} \phi \partial_{\nu} \phi : g^{\mu\nu}
\sqrt{-g} - \frac1{4!} \lambda : \phi^4 : \sqrt{-g} + {\Delta {\cal L}} \;.
\end{equation}
The colons stand for a covariant normal-ordering prescription whose effect
is to subtract off contributions from coincident propagators \cite{AbWo1}.
The counterterms ${\Delta {\cal L}}$ are used to enforce masslessness and
the vanishing of the stress-energy tensor at the initial time. We are not 
quantizing gravity. The metric is a non-dynamical background which we take 
to be locally de Sitter in conformal coordinates,
\begin{equation}
g_{\mu \nu}(\eta,\vec{x}) = \Omega^2(\eta) \eta_{\mu\nu} \qquad , \qquad
\Omega(\eta) = -{1 \over H \eta} = e^{Ht} \;.
\end{equation}
(Recall that $\Lambda = 3 H^2$.) To regulate the infrared problem on
the initial value surface we work on the manifold $T^3 \times R$, with the
spatial coordinates in the finite range, $-H^{-1}/2 < x^i \leq H^{-1}/2$.
We release the state in Bunch-Davies vacuum at $t = 0$, corresponding to
conformal time $\eta = -H^{-1}$. Note that the infinite future corresponds
to $\eta \rightarrow 0^-$, so the possible variation of conformal
coordinates in either space or time is at most ${\Delta x} = {\Delta \eta}
= H^{-1}$.

The stress-energy tensor consists of the potential term plus derivatives
which are subdominant \cite{TsWo3},
\begin{equation}
T_{\mu\nu}(x) = - g_{\mu\nu}(x) \frac{\lambda}{4!} : \phi^4(x) : + \; {\rm 
subdominant} \; .
\end{equation}
Because there are no coincident propagators the lowest contribution to
its expectation value comes at order $\lambda^2$ from the two diagrams in 
Fig.~1 \cite{TsWo3}. (Of course there is a cosmological counterterm to absorb 
the ultraviolet divergence.) Further, the derivatives on the top vertex of 
the right hand diagram render its contribution subdominant to the left hand 
diagram in powers of $Ht$. So the dominant contribution to the expectation 
value of the stress-energy tensor is $- g_{\mu\nu}(x)$ times,
\begin{eqnarray}
\lefteqn{\frac{\lambda}{4!} \left\langle \Omega \left\vert : \phi^4(x) : 
\right\vert \Omega \right\rangle =} \nonumber \\
& & \frac{-i \lambda^2}{4!} \int_{t'>0} d^4x' \Omega^4(\eta') \left\{ \left[ 
{i\Delta}_{++}(x;x') \right]^4 - \left[ {i\Delta}_{+-}(x;x') \right]^4 \right\}
+ O(\lambda^3) \; . \label{phi4}
\end{eqnarray}
The difference of $++$ and $+-$ propagators,\footnote{The conformal coordinate 
separations in these formulae are, ${\Delta x} \equiv \Vert \vec{x} - 
\vec{x}'\Vert$ and ${\Delta \eta} \equiv \eta - \eta'$.}
\begin{eqnarray}
\lefteqn{{i\Delta}_{++}(x;x') \equiv } \nonumber \\
& & {1 \over 4 \pi^2} {\Omega^{-1}(\eta) \Omega^{-1}(\eta') \over 
{\Delta x}^2 - (\vert {\Delta \eta} \vert - i \epsilon)^2} - {H^2 \over 
8 \pi^2} \ln\left[ H^2 \left( {\Delta x}^2 - (\vert {\Delta \eta} \vert 
- i \epsilon)^2 \right) \right] \; , \\
\lefteqn{{i\Delta}_{+-}(x;x') \equiv } \nonumber \\
& & {1 \over 4 \pi^2} {\Omega^{-1}(\eta) \Omega^{-1}(\eta') \over 
{\Delta x}^2 - ({\Delta \eta} + i \epsilon)^2} - {H^2 \over 8 \pi^2} 
\ln\left[ H^2 \left( {\Delta x}^2 - ({\Delta \eta} + i \epsilon)^2 
\right) \right] \; ,
\end{eqnarray}
comes from using the Schwinger-Keldysh formalism \cite{Schwinger,Jordan} to 
compute an expectation value rather than an in-out amplitude. This form 
ensures that the result is real and that it depends only upon points 
$x^{\prime \mu}$ in the past light cone of the observation point $x^{\mu}$. 
The lower limit of temporal integration at $\eta' = -H^{-1}$ (that is, $t' 
= 0$) derives from the fact that we release the state in free Bunch-Davies 
vacuum at this instant.

Although (\ref{phi4}) was computed in ref.~\cite{TsWo3} we will go over it 
in detail. Since only the logarithm term of the propagator breaks conformal 
invariance it is perhaps not surprising that the dominant secular effect 
comes from taking this term in each of the four propagators. This contribution 
is completely ultraviolet finite, and its evaluation is straightforward if 
one goes after only the largest number of temporal logarithms,
\begin{eqnarray}
\lefteqn{\frac{-i \lambda^2}{4!} \left({-H^2 \over 8 \pi^2}\right)^4 
\int_{-H^{-1}}^{\eta} d\eta' \left({-1 \over H \eta'}\right)^4 4\pi 
\int_0^{\infty} dr r^2} \nonumber \\
& & \qquad \times \left\{ \ln^4\left[ H^2 \left( r^2 - ({\Delta \eta} -i
\epsilon)^2\right) \right] - \ln^4\left[ H^2 \left( r^2 - ({\Delta \eta} +i
\epsilon)^2\right) \right] \right\} \nonumber \\
& & \rightarrow {-i \lambda^2 H^4 \over 2^{13} 3^1 \pi^7} \int_{-H^{-1}}^{\eta}
d\eta' {1 \over \eta^{\prime 4}} \int_0^{\Delta \eta} dr r^2 \; 8\pi i \;
\ln^3\left[H^2 ({\Delta \eta}^2 - r^2)\right] \; , \\
& & = {\lambda^2 H^4 \over 2^{10} 3^1 \pi^6} \int_{-H^{-1}}^{\eta} d\eta' 
{{\Delta \eta}^3 \over \eta^{\prime 4}} \int_0^1 dx x^2 \left[ 2 \ln(H 
{\Delta \eta}) + \ln(1 - x^2)\right]^3 \; , \\
& & \rightarrow {\lambda^2 H^4 \over 2^7 3^2 \pi^6} \int_{-H^{-1}}^{\eta}
d\eta' {{\Delta \eta}^3 \over \eta^{\prime 4}} \ln^3(H {\Delta \eta}) \; .
\end{eqnarray}
For large $Ht$ the biggest effect comes from the term with the most factors
of $\ln(-H\eta) = -Ht$. That the integrand contributes three such factors
follows from the expansion,
\begin{equation}
\ln(H{\Delta \eta}) = \ln(-H \eta') - \sum_{n=1}^{\infty} \frac1{n} \left({
\eta \over \eta'}\right)^n \; .
\end{equation}
An additional factor comes from performing the integration up against the 
final term in the expansion of the ratio,
\begin{equation}
{{\Delta \eta}^3 \over \eta^{\prime 4}} = {\eta^3 \over \eta^{\prime 4}} - 3
{\eta^2 \over \eta^{\prime 3}} + 3 {\eta \over \eta^{\prime 2}} - \frac1{
\eta'} \; .
\end{equation}
The final result is therefore,
\begin{equation}
\frac{\lambda}{4!} \left\langle \Omega \left\vert : \phi^4(x) : \right\vert
\Omega \right\rangle = - {\lambda^2 H^4 \over 2^9 3^2 \pi^6} \left\{ (Ht)^4
+ O(H^3 t^3) \right\} + O(\lambda^3) \; . \label{final}
\end{equation}

Two points deserve comment. First, there is nothing paradoxical about the 
negative sign of the $(Ht)^4$ contribution to the expectation value of a 
positive operator. The actual result is dominated by a positive ultraviolet 
divergent constant. It is only after the cosmological counterterm is used to 
subtract this divergence that the ultraviolet finite factor of $(Ht)^4$ 
dominates the late time behavior of the scalar stress-energy tensor at order 
$\lambda^2$. 

Our second comment is that the negative sign of the $(Ht)^4$ term has a 
simple physical interpretation. As the inflationary expansion rips more and 
more scalars out of the vacuum their attractive self-interaction acts to 
pull them back together. The resulting expansion rate is \cite{TsWo3},
\begin{equation}
\dot{b}(t) = H \left\{1 - {\lambda^2 G \Lambda \over 2^7 3^4 \pi^5} \left[
(Ht)^4 + O(H^3 t^3)\right] + O(\lambda^3,G^2)\right\} \; . \label{scalarH}
\end{equation}
This is the direct analog of the graviton effect we have been seeking. We
turn now to the problem of computing the left hand diagram of Fig.~1 in the 
presence of the back-reacted geometry.

\section{A new infrared expansion}

It is straightforward to work out the interaction vertices in any background.
The great obstacle to computing in different backgrounds is finding the 
propagator. We need to do this in a homogeneous, isotropic and spatially flat 
background,
\begin{equation}
g_{\mu\nu} dx^{\mu} dx^{\nu} = -dt^2 + e^{2 b(t)} d{\vec x} \cdot d{\vec x}
= \Omega^2(\eta) \Bigl( -d\eta^2 + d{\vec x} \cdot d{\vec x}\Bigr) \; , 
\end{equation}
which began de Sitter inflation at $\eta = -1/H$ in free Bunch-Davies vacuum. 
The problem here is solving for $u(\eta,k)$, the mode function for a scalar of
co-moving wave number $k$. It obeys the equation,
\begin{equation}
u^{\prime\prime}(\eta,k) + 2 \frac{\Omega'}{\Omega} u^{\prime}(\eta,k)
+ k^2 u(\eta,k) = 0 \; . \label{modeeqn}
\end{equation}
Canonical quantization and the initial Bunch-Davies vacuum fixes the
initial conditions (up to a phase) as,
\begin{equation}
u\left(-H^{-1},k\right) = {i H \over \sqrt{2 k^3}} e^{ik/H} \left\{ 1 - i
\frac{k}{H} \right\} \quad , \quad u'\left(-H^{-1},k\right) = { i H \over
\sqrt{2 k^3}} e^{ik/H} \left\{ -\frac{k^2}{H} \right\} \; . \label{canonical}
\end{equation}

If we could solve (\ref{modeeqn}) for arbitrary conformal factor $\Omega(\eta)$
the $++$ and $+-$ propagators would be,
\begin{eqnarray}
{i\Delta}_{++}(x;x') & = & \int {d^3k \over (2\pi)^3} e^{i\vec{k} \cdot 
{\Delta\vec{x}}} \Bigl\{\theta({\Delta \eta}) u(\eta,k) u^*(\eta',k) + 
\theta( - {\Delta \eta}) u^*(\eta,k) u(\eta',k)\Bigr\} \; , \label{++prop} \\
{i\Delta}_{+-}(x;x') & = & \int {d^3k \over (2\pi)^3} e^{i\vec{k} \cdot 
{\Delta\vec{x}}} u^*(\eta,k) u(\eta',k)\Bigr\} \; . \label{+-prop}
\end{eqnarray}
Here we define the conformal coordinate separations as, ${\Delta\vec{x}} \equiv 
\vec{x} - \vec{x}'$ and ${\Delta \eta} \equiv \eta - \eta'$. A closely related 
and often useful quantity is the retarded Green's function,
\begin{equation}
G_{\rm ret}(x;x') = \theta({\Delta \eta}) \int {d^3k \over (2\pi)^3} e^{i 
\vec{k} \cdot {\Delta\vec{x}}} {\rm Im}\Bigl\{ u(\eta,k) u^*(\eta',k) 
\Bigr\} \; . \label{Greens}
\end{equation}

Since the physical effect we are studying is infrared we actually require only
an expansion which is valid for long wavelengths. What this means changes with
time owing to the expansion of spacetime. During inflation we distinguish
between modes which infrared (IR) and ultraviolet (UV) by comparing their 
co-moving wavenumbers thusly,
\begin{equation}
H < {\rm IR} < {\Omega'(\eta) \over \Omega(\eta)} < {\rm UV} \; .
\end{equation}
Suppose inflation ends at conformal time $\eta_1$, and that subsequent 
evolution consists either of deflation or subluminal expansion. Then some of 
the IR modes which experienced horizon crossing during inflation re-enter the
horizon. We designate these modes as thermal (TH) and distinguish them thusly,
\begin{equation}
H < {\rm IR} < \left\vert{\Omega'(\eta) \over \Omega(\eta)}\right\vert < {\rm
TH} < {\Omega'(\eta_1) \over \Omega(\eta_1)} < {\rm UV} \; .
\end{equation}

Since we need an infrared expansion the straightforward approach would be to
solve (\ref{modeeqn}) perturbatively, regarding the $k^2$ term as the small
parameter. This turns out not to give a satisfactory expansion for a number of 
reasons \cite{Unruh,AbWo2}. One problem is that the initial conditions involve
fractional, and even inverse powers of the small parameter. Another problem 
is that although causality requires the full retarded Green's function 
(\ref{Greens}) to vanish for ${\Delta x} > {\Delta \eta}$, this feature is not 
manifest at any finite order of the $k^2$ expansion. A better strategy seems 
to be first factoring out a temporal exponential and a multiplicative
normalization,
\begin{equation}
u(\eta,k) \equiv {i H \over \sqrt{2 k^3}} e^{-i k \eta} q(\eta,k) \; .
\end{equation}
We then perturb, regarding $ik$ as the small parameter,
\begin{equation}
q''(\eta,k) + 2 {\Omega' \over \Omega} q'(\eta,k) = 2 i k \left\{ q'(\eta,k)
+ {\Omega' \over \Omega} q(\eta,k)\right\} \; .
\end{equation}
The initial conditions (\ref{canonical}) become,
\begin{equation}
q\left(-H^{-1},k\right) = 1 - i \frac{k}{H} \qquad , \qquad 
q'\left(-H^{-1},k\right) = i k \; .
\end{equation}

Making the substitution,
\begin{equation}
q(\eta,k) = \sum_{n=0}^{\infty} q_n(\eta) (ik)^n \; ,
\end{equation}
we easily find the first two terms,
\begin{equation}
q_0(\eta) = 1 \qquad , \qquad q_1(\eta) = \eta \; .
\end{equation}
The next order term is,
\begin{equation}
q_2(\eta) \equiv \frac12 \Bigl(\eta^2 - H^{-2}\Bigr) + H^{-1} \int_{-1/H}^{
\eta} d\eta' \Omega^{-2}(\eta') + \int_{-1/H}^{\eta} d\eta' \Omega^{-2}(\eta')
\int_{-1/H}^{\eta'} d\eta'' \Omega^2(\eta'') \; ,
\end{equation}
and the higher terms are obtained through the recursion relation,
\begin{equation}
q_n(\eta) \equiv \int_{-1/H}^{\eta} d\eta' q_{n-1}(\eta') + \int_{-1/H}^{
\eta} d\eta' \Omega^{-2}(\eta') \int_{-1/H}^{\eta'} d\eta'' \Omega^2(\eta'')
q'_{n-1}(\eta'') \; .
\end{equation}

This expansion avoids the problems associated with the long wavelength 
limit \cite{Unruh,AbWo2}. It has the potential for extending to the 
ultraviolet regime by virtue of the oscillatory behavior of the factor of 
$e^{-ik\eta}$. The first two terms are also exact for pure de Sitter, and 
the higher terms can be evaluated in the slow roll expansion. Finally, 
applying the expansion to the mode sum for the retarded Green's function 
results in a series of higher and higher derivatives of the past light 
cone theta function $\theta({\Delta \eta} - {\Delta x})$,
\begin{equation}
G_{\rm ret}(x;x') = -\theta({\Delta \eta}) {H^2 \over 4\pi} \left\{
\theta({\Delta \eta} - {\Delta x}) + {\eta \eta' \over {\Delta x}} \delta(
{\Delta \eta} - {\Delta x}) + \dots \right\} \; .
\end{equation}
This certainly seems to be organized as an infrared expansion should be, with 
the large volume effects at lowest order.

\section{Post-inflationary deflation}

The cosmological constant is screened by the coherent superposition of the
attractive long range potentials produced by virtual particles which are 
ripped apart during inflation. Although any one particle creation event 
makes only a small contribution, the coherent superposition of events from 
the entire past light cone can be enormous. Two things happen when the 
expansion rate begins to slow appreciably: the volume rate of particle 
production drops and the growth of the past light cone increases. The first
can not have much immediate effect on screening because most of the past 
light cone at the end of inflation derives from periods when the particle
production rate was still high. It is the second effect which dominates.
Slower expansion means that more and more invariant volume from the period
of inflationary particle production becomes visible. Further, most of the
propagation of this effect occurs through the period of rapid expansion
when the de Sitter approximation is still valid. Not only does screening 
continue to grow after the end of inflation; its rate of growth actually
{\it increases}.

The situation I have just described is a classic runaway instability. The 
lower the expansion rate drops the faster the attractive self-interaction 
accumulates. There seems to be no alternative to a rapid decay into 
deflation. To gain a rough understanding of what happens next let us 
solve for the mode functions under the assumption that pure de Sitter
inflation goes over to pure de Sitter deflation at conformal time $\eta_1$,
\begin{equation}
\Omega(\eta) = \cases{-1/{H\eta} , & $-H^{-1} \leq \eta \leq \eta_1$;\cr
1/H(\eta - 2\eta_1) , & $\eta_1 \leq \eta$ .\cr}
\end{equation}
The inflationary mode function was given in the previous section,
\begin{equation}
u(\eta,k) = {i H \over \sqrt{2 k^3}} (1 + i k \eta) e^{-i k \eta} \qquad 
\forall \; {-H}^{-1} \leq \eta \leq \eta_1 \; .
\end{equation}
Since the conformal factor is continuous at $\eta = \eta_1$ the mode function
is also continuous there, as is its first derivative. The deflationary mode 
function can be most simply expressed in terms of another function $v(\eta,k)$,
\begin{eqnarray}
u(\eta,k) & = & -v(\eta,k) + {i e^{-i k \eta_1} \over k \eta_1} \left[ e^{i k
\eta_1} v(\eta,k) + e^{-i k \eta_1} v^*(\eta,k)\right] \qquad \forall \; 
\eta_1 \leq \eta \; , \label{defmode} \\
v(\eta,k) & = & {i H \over \sqrt{2 k^3}} \Bigl( 1 + i k (\eta - 2 \eta_1)
\Bigr) e^{-i k \eta} \; .
\end{eqnarray}
Note the factor of $1/\eta_1$ on the second term in (\ref{defmode}). Since
screening requires an enormous number of e-foldings to become effective, we
can assume that $\eta_1$ is an infinitesimal negative number. It follows that
the second term of (\ref{defmode}) rapidly becomes huge. Since this appears
to make screening {\it even stronger} there does not seem to be any restoring
force against runaway deflation!

Appearances are deceiving. The growth in the second term of (\ref{defmode})
has a physical origin which becomes clear upon simplification,
\begin{equation}
u(\eta,k) = -v(\eta,k) + i e^{-i k \eta_1} \sqrt{2 \over H} \left({H \over k}
\right)^{\frac32} {\Omega_1 \over \Omega(\eta)} \left\{ \cos\left[k (\eta - 
\eta_1)\right] - {\sin[k (\eta - \eta_1)] \over k (\eta - 2\eta_1)} \right\} 
\; . \label{decomp}
\end{equation}
Now specialize the classification of co-moving wave numbers to the deflationary
geometry,
\begin{equation}
H < {\rm IR} < {1 \over \eta - 2 \eta_1} < {\rm TH} < {-1 \over \eta_1} <
{\rm UV} \; .
\end{equation}
After any significant amount of deflation we have $\eta \gg -\eta_1$. We can
therefore conclude that most infrared modes obey $k \eta \ll 1$ and hence,
\begin{equation}
\cos\left[k (\eta - \eta_1)\right] - {\sin[k (\eta - \eta_1)] \over k (\eta 
- 2\eta_1)} \approx -\frac13 (k \eta)^2 \; .
\end{equation}
The second term is therefore not large for modes which are still infrared. 
However, most of the infrared modes which have re-entered the horizon --- 
which we call ``thermal'' --- obey $1 \ll k\eta$. For them the cosine term 
is of order one whereas the sine is minuscule. The result is that the second 
term of (\ref{decomp}) grows exponentially like the inverse scale factor,
\begin{equation}
{\Omega_1 \over \Omega(\eta)} = e^{H (t - t_1)} \; .
\end{equation}
What we are seeing in the second term is nothing more than the blue shift 
due to deflation. The thermal modes have been populated by the particle
production that went on during inflation, but after re-entering the horizon
they obey the equation of state of radiation. As the universe deflates they
form a hot, dense plasma.

It is not correct to treat this second term as part of the mode function 
for a quantum field. It is rather a stochastic background whose precise value
is random but definite in the sense advocated by Linde \cite{Linde,AbWo2}. To
understand this let us form the Fourier transform of the free field by 
multiplying the mode functions by canonically normalized creation and 
annihilation operators,
\begin{equation}
\widetilde{\phi}(\eta,\vec{k}) = u(\eta,k) a(\vec{k}) + u^*(\eta,k)
a^{\dagger}(\vec{k}) \; .
\end{equation}
This field commutes canonically with $\Omega^2$ times it conformal time
derivative. Suppose we decompose the field as we have the mode function
(\ref{decomp}). This results in two terms,
\begin{eqnarray} 
\widetilde{\phi}_{\rm qm}(\eta,\vec{k}) & \equiv & -v(\eta,k) a(\vec{k}) - 
v^*(\eta,k) a^{\dagger}(\vec{k}) \; , \\
\widetilde{\phi}_{\rm big}(\eta,\vec{k}) & \equiv & \sqrt{2 \over H} 
\left({H \over k}\right)^{\frac32} {\Omega_1 \over \Omega(\eta)} \left\{ 
\cos\left[k (\eta - \eta_1)\right] - {\sin[k (\eta - \eta_1)] \over k 
(\eta - 2\eta_1)} \right\} \nonumber \\
& & \hspace{5cm} \times \left[i e^{-i k \eta_1} a(\vec{k}) - i e^{i k \eta_1} 
a^{\dagger}(\vec{k})\right] \; .
\end{eqnarray}
The first of these, $\widetilde{\phi}_{\rm qm}$ obeys the same commutation
relations as the full field. However, $\widetilde{\phi}_{\rm big}$ {\it
commutes} with its conformal time derivative. It is still an operator, in
that it has the potential for taking any value, but it behaves like a classical
field once this value is chosen. 

One must subsume $\widetilde{\phi}_{\rm big}(\eta,\vec{k})$ into the 
background. As we saw, it is only significant for the thermal modes. These 
terms should be treated as a classical gas. Doing this permits us to keep 
using perturbation theory. Far more important, it also provides the physical 
mechanism by which deflation is halted. For note that screening is maintained 
by the propagation into the future of quantum correlations from the period of 
inflationary particle production. This is accomplished by the infrared modes. 
But they will be scattered by the thermal particles, and such a collision is 
overwhelmingly likely to bump them up into the high momentum regime. We can 
therefore view the thermal modes as forming a partial barrier to the forward 
propagation of the infrared modes. As the universe deflates the barrier 
becomes thicker and thicker. Eventually it overcomes the continued growth of
the past light cone, at which point the bare cosmological constant begins to 
reassert itself and expansion resumes.

Computing the scattering rate, $\Gamma(\eta,k)$, is a trivial exercise in 
perturbation theory and has already been done under the assumption that the 
previous expansion for $u(\eta,k)$ is valid in the absence of the barrier. The 
depletion effect can be roughly accounted for by degrading the infrared mode 
functions by the following replacement,
\begin{equation}
u(\eta,k) \longrightarrow \exp\left[-\frac12 \int^{\eta} d\eta' 
\Gamma(\eta',k)\right] \times u(\eta,k) \; .
\end{equation}
Simple phase space arguments imply that $\Gamma(\eta,k)$ is proportional to 
$1/k$. The wonderful feature of this result is that the mode sum for the 
propagator can be evaluated analytically using the integral,
\begin{equation}
\int_0^{\infty} dk k^{\nu-1} \exp\left[-\frac{\beta}{k} - \gamma k\right] =
2 \left(\frac{\beta}{\gamma}\right)^{\frac{\nu}2} K_{\nu}\left(2 \sqrt{\beta 
\gamma}\right) \qquad {\rm Re}(\beta) > 0 \; , \; {\rm Re}(\gamma) > 0 
\; .
\end{equation}
So it should be possible to evaluate the strength of the quantum effect as the
geometry back-reacts.

\section{Subsequent evolution}

Post inflationary evolution is controlled by the balance between the 
degradation of the screening effect by the thermal barrier, and the fact that
more and more of the inflationary past light cone is visible at later and
later times. The key to stability is the thermal barrier. If the expansion
rate becomes too high the density of particles in the barrier decreases, which
reduces the new scatters of infrared quanta and therefore makes screening more
effective. If the expansion rate slows too much the particle density falls off
slower, leading to more new scatters and a decline in screening. Note that the
growth of the past light cone has a destabilizing effect since it grows more
slowly when the expansion rate is higher. What gives rise to stability is that
the thermal degradation effect acts more strongly and more quickly.

The growth of the past light cone is a purely geometrical effect which can be 
usefully approximated as follows. Suppose that inflation ends suddenly at 
conformal time $\eta_1$ (co-moving time $t_1$). Then the invariant volume of 
the inflationary
past light cone which is visible at some later conformal time $\eta$ (co-moving
time $t$) is,
\begin{equation}
V(t) = \frac43 \pi \int_{-1/H}^{\eta_1} d\eta' \Omega^4(\eta')
(\eta - \eta')^3 \approx \frac43 \pi \left\{ H t_1 + I^3(t) \right\} \; ,
\label{plc}
\end{equation}
where the function $I(t)$ is,
\begin{equation}
I(t) \equiv H e^{b(t_1)} \int_{t_1}^t dt' e^{-b(t')} \; .
\end{equation}
Because the universe deflates immediately after inflation $b(t_1) > b(t)$,
so $I(t)$ rapidly comes to dominate the constant term in (\ref{plc}).

The nature of the thermal barrier will of course depend upon the available
matter quanta into which energy flows through thermalization. Combining 
the various effects should result in a nonlocal equation for the logarithmic 
scale factor $b(t)$ which can be evolved numerically. Questions to be answered 
include:
\begin{enumerate}
\item{What reheat temperature is reached?}
\item{What is the asymptotic form of $b(t)$?}
\item{What is the impact on structure formation?}
\item{How does the model respond to late time phase transitions?}
\item{Does the model possess a late time phase of acceleration such as
seems to be occurring now \cite{Reiss,Perl}?}
\end{enumerate}

\Acknowledgements

I have profited from years of collaboration and friendship with L. R. Abramo
and N. C. Tsamis. I also thank J. Garcia-Bellido, A. Liddle and M. Sasaki for
enjoyable and informative discussions. This work was partially supported by 
DOE contract DE-FG02-97ER\-41029 and by the Institute for Fundamental Theory.

\end{document}